\documentclass[aps,prl,twocolumn,twoside,superscriptaddress,nofootinbib,preprintnumbers]{revtex4-1}

\usepackage[bookmarks=false]{hyperref}
\usepackage{graphicx}
\usepackage{amsmath}
\usepackage{xspace}
\usepackage{color}
\usepackage[normalem]{ulem}

\newcommand{\eq}[1]{Eq.~\eqref{eq:#1}}
\newcommand{\eqs}[2]{Eqs.~\eqref{eq:#1} and \eqref{eq:#2}}

\newcommand{\df}{\mathrm{d}}

\newcommand{\tr}{\textrm{tr}}

\newcommand{\al}{\alpha}

\newcommand{\nn}{\nonumber}

\def\zcut{z_{\text{c}}}
\def\Rcut{R_{c}}
\def\Ecut{E_{c}}
\def\Rmax{R}

\allowdisplaybreaks[2]

\begin{document}


\preprint{\vbox{\hbox{Nikhef 2018-052}}}

\title{The Entropy of a Jet}

\author{Duff Neill}

\affiliation{Theoretical Division, MS B283, Los Alamos National Laboratory, Los Alamos, NM 87545, USA\vspace{0.5ex}}

\author{Wouter J.~Waalewijn}

\affiliation{Institute for Theoretical Physics Amsterdam and Delta Institute for Theoretical Physics, University of Amsterdam, Science Park 904, 1098 XH Amsterdam, The Netherlands}
\affiliation{Nikhef, Theory Group, Science Park 105, 1098 XG, Amsterdam, The Netherlands}

\begin{abstract}
Scattering processes often inevitably include the production of infrared states, which are highly correlated with the hard scattering event, and decohere the hard states. This can be described using the entropy of the hard reduced density matrix, which is obtained from tracing over infrared states. We determine this entropy for an asymptotically-free gauge theory by separating the Hilbert space into hard and infrared states, and calculate it in a leading-logarithmic approximation for jets. We find that the entropy increases when the resolution scales defining the hard radiation are lowered, that this entropy is related to the subjet multiplicity, and explore connections to using jet images for machine learning, and the forward-scattering density matrix of partons in a nucleon probed in deep-inelastic scattering.
\end{abstract}

\maketitle

{\it Introduction. --}
In scattering experiments, one studies the short-distance collision and initial state through the final-state remnants. The emission of arbitrarily soft quanta, as found in electrodynamics or gravity, limits how well one can constrain the final state (an issue that may have connections to the blackhole information paradox, see \cite{Hawking:2016msc,Haco:2018ske}). In a confining gauge theory, the mass gap curtails soft particle production. However, an asymptotically-free gauge theory probed at sufficient energy will have soft and collinear production of particles throughout the regions of spacetime transversed by hard partons, well before the mass gap is reached, leading to jets. Practically, this implies a similar limitation about how much one can learn concerning a scattering event, prompting the question: How is the information about the hard scattering or the initial state distributed over the scattering remnants, and how much information is lost in generating these infrared states? (Ref.~\cite{Berges:2017hne} considered entropy production in a gauge theory using a non-perturbative fragmentation model, similar to the Lund string model~\cite{Andersson:1983ia,Andersson:1998tv}. Our discussion will be strictly perturbative, seeking to capture the entropy produced in the partonic cascade before any non-perturbative fragmentation.)

In this Letter, we quantify the correlation between the hard parton produced in the scattering and the soft and collinear radiation, by calculating the entropy of the hard reduced density matrix of a single jet, in a leading logarithmic (LL) approximation.

The correlation of soft and collinear radiation with the scattering event is also an active area of research in quantum chromodynamics (QCD), and is experimentally tested at the Large Hadron Collider and the Relativistic Heavy Ion Collider. For example, the field of jet substructure aims to deduce the quantum numbers of the particles in some hard interaction, from their soft and collinear emissions (for a review, see \cite{Larkoski:2017jix}). Likewise, in deep-inelastic scattering, the struck parton occupies a small region of spacetime while it interacts with the exchanged photon, but is entangled with the gluon and quark fields that permeate the proton. In both cases a significant amount of this correlation is encoded by perturbative collinear and soft splittings, and within jet substructure various studies have attempted to quantify this~\cite{Larkoski:2014pca,Datta:2017rhs}.

Interestingly, these examples are formally connected, as a conformal transformation relates the structure of soft emissions in the final state of a hard scattering process~\cite{Dasgupta:2001sh,Banfi:2002hw,Weigert:2003mm} to initial-state bremsstralung of soft radiation~\cite{McLerran:1993ka,McLerran:1993ni,McLerran:1994vd,Balitsky:1995ub,Kovchegov:1999yj,JalilianMarian:1996xn,JalilianMarian:1997gr,Iancu:2001ad}, as explored in \cite{Hatta:2008st,Avsar:2009yb,Caron-Huot:2015bja}. Thus much of our discussion (since to LL accuracy we are dealing with collinear \emph{and} soft limits) should have an analog within forward-scattering physics, where entropy production in soft initial-state radiation has been considered~\cite{Kutak:2011rb,Kovner:2015hga,Kharzeev:2017qzs,Hagiwara:2017uaz,Kovner:2018rbf}. This literature has reached differing conclusions about the amount of entropy associated with a parton in a nucleon, which are compared to our results. Entropy production due to correlations in momentum space has been considered in \cite{Balasubramanian:2011wt,Hsu:2012gk}, while \cite{Lello:2013bva} examined entropy due to tracing over unobserved products in a particle decay.

{\it Hard reduced density matrix. --}
The hard reduced density matrix results from (roughly) tracing over all soft radiation below an energy scale $\Ecut$ and all collinear radiation below an angular scale $\Rcut$. It can be used to formulate a Monte Carlo parton shower tracking color and spin coherence effects~\cite{Nagy:2007ty,Nagy:2014mqa}. We will calculate its entropy using all $n$-subjet \emph{differential} cross sections, where a jet is decomposed into subjets with opening angle $\Rcut$ and minimum energy $\Ecut$. (Analogously, jets produced in a hard scattering can be considered, instead of subjets in jets.) As we will discuss, these cross sections correspond to diagonal terms in the reduced density matrix, whose rows and columns are indexed by the number of subjets and their momenta. This involves dividing the phase space 
into resolved and unresolved regions (the subjets and their interiors), using for instance the jet algorithm in \cite{Ellis:1993tq,Dokshitzer:1997in}, or more formally with the stress-energy tensor \cite{Sveshnikov:1995vi,Tkachov:1995kk,Tkachov:1999py}. 

We may think of a subjet to be a hard or resolved ``state'' that is dressed by further soft and collinear emissions. In the study of the factorization of amplitudes (see e.g.~\cite{Feige:2014wja,Larkoski:2014bxa}), the hard state is approximated by a specific on-shell partonic state. The soft and collinear emissions below the scale $E_c$ and $\Rcut$ will \emph{decohere} various quantum numbers of these hard states above the scale $E_c$ and $\Rcut$. For instance, superpositions of distinct momentum states would be destroyed, thus selecting a specific basis that diagonalizes the hard reduced density matrix, as argued in \cite{PhysRevA.63.032102,PhysRevA.67.042702,Carney:2017jut}. A medium can alter this decoherence process, which is important for jets propagating through a heavy-ion collision~\cite{Mehtar-Tani:2013pia}. The diagonal terms in this basis represent the quasi-classical probability density to observe that basis state. 

Though we focus on the terms that are diagonal in the number $n$ of hard emissions (subjets), the reduced density matrix element, defined on hard states, has the general form: 
\begin{align}\label{eq:reduced_density_matrix_ELEMENT}
&\rho_{n}\bigl(\{p_i\}_{i=1}^n,\{p_i^{\:\prime}\}_{j=1}^m\bigr)
\nn \\ &\quad
= \sum_{\{a_i,\lambda_i,f_i\}_{i=1}^n}\sum_{\{a_j',\lambda_j',f_j'\}_{j=1}^m}C_H^{\dagger}\bigl(p_1^{a_1\lambda_1f_1},...,p_n^{a_n\lambda_nf_n}\bigr)
\nn \\ & \qquad \times
I\bigl(p_1^{a_1\lambda_1f_1},...,p_n^{a_n\lambda_nf_n};p_1'\,^{a_1'\lambda'_1f_1'},...,p_m'\,^{a_m'\lambda'_mf_m'}\bigr)
\nn \\ & \qquad \times
C_H\bigl(p_1'\,^{\,a_1'\lambda'_1f_1'},...,p_m'\,^{a_m'\lambda'_mf_m'}\bigr)+...
\end{align}
Here $p_i'$ denotes the momentum, $a_i'$ the color, $\lambda_i'$ the spin and $f_i'$ the flavor of particle $i$ in the hard amplitude $C_H$, with the corresponding unprimed variables for the conjugate amplitude $C_H^\dagger$. The function $I$ is a \emph{combination} of functions similar to the soft functions (matrix element of eikonal Wilson lines) and collinear functions found in factorization using Soft-Collinear Effective Theory~\cite{Bauer:2000yr,Bauer:2001yt,Bauer:2001ct} (see e.g.~\cite{Ellis:2010rwa,Stewart:2010tn} for examples of infrared functions for exclusive $n$-jet cross sections, and~\cite{Larkoski:2015kga} for extensions to subjets) or in the Collins-Soper-Sterman (CSS) approach (see e.g.~\cite{Collins:1981uk,Kidonakis:1998nf}). It describes production of the soft and collinear emissions below the scale $E_c$ and $R_c$ that we used to define the hard radiation, and is related to the Feynman-Vernon influence functional in the decoherence literature~\cite{OpenSysBook}. We have
\begin{align}\label{eq:momentum_decoherence}
&I\bigl(p_1^{a_1\lambda_1f_1},...,p_n^{a_n\lambda_nf_n};p_1'\,^{a_1'\lambda'_1f_1'},...,p_m'\,^{a_m'\lambda'_mf_m'}\bigr)=0\,, 
\nn \\ &\text{unless } n = m, p_i = p_i' \text{ and } a_i=a_i' \text{ for all } i,
\end{align}
because otherwise infrared divergences do not cancel. (For a discussion in the context of heavy non-relativistic particles coupled to photons, see~\cite{OpenSysBook}. A calculation of the influence functional for distinct spacetime paths in the amplitude and conjugate amplitude reveals that, at late times, infrared divergences drive it to zero.)

Specifically, soft radiation forces the directions of the momenta and the gauge representations of the particles (representing the subjets) to be equal in the amplitude and conjugate amplitude, while collinear radiation forces the energies to be equal. Focusing on QCD, the SU(3) representation separates quarks from gluons. We will assume that quark flavors and spins are not observed, tracing over them.

Consequently, the diagonal reduced density matrix is given by the following sum over exclusive $n$-subjet cross sections: 
\begin{align}\label{eq:reduced_density_matrix}
\rho&=\sum_{n=1}^{\infty}\int_H \df \Pi_n(p_J)\frac{1}{\sigma} \frac{\df\sigma}{\df\Pi_n}|p_1,p_2,...p_n\rangle\langle p_1,p_2,...p_n|
\,,\nn\\
\df \Pi_n(p_J) &=\Big(\prod_{i=1}^n\frac{\df^4p_i}{(2\pi)^3}\,\delta(p_i^2)\Big)\,\delta^{4}\Big(p_J - \sum_{i=1}^np_i\Big)\,.
\end{align}
Here $\df \Pi_n$ denotes the on-shell phase space of $n$-hard emissions, $p_J^\mu$ the jet momentum, and the sum starts at $n=1$ corresponding to one subjet. The integral is restricted to the hard region of phase space, denoted by the subscript $H$. We have normalized the differential cross section, so that we may interpret the differential cross section as a probability density $P$, i.e.~$\rho_{n}\bigl(\{p_i\}_{i=1}^n,\{p_i\}_{i=1}^n\bigr) = ({1}/{\sigma})\, {\df\sigma}/{\df\Pi_n}= {\df P}/{\df \Pi_n}$.

{\it Entropy. --}
We first consider the Renyi entropy of the reduced density matrix: 
\begin{align}\label{eq:Renyi_Entropy}
  \mathcal{S}_\alpha&=\frac{\text{ln tr}[\rho^\alpha]}{1-\alpha}=\frac{1}{1-\alpha}\,\ln\sum_{n=0}^{\infty}\tr[\rho_{n}^{\alpha}]\,,\nn \\
  \text{tr} [\rho_{n}^{\alpha}] & =\!\int_H {\rm d} \Pi_n\,\bigl[\rho_{n}\bigl(\{p_i\}_{i=1}^n,\{p_i\}_{i=1}^n\bigr)\bigr]^\alpha\,.
\end{align}
We can now obtain the von Neumann entropy by taking the limit $\alpha\rightarrow 1$, 
\begin{align}\label{eq:von_neumann_entropy}
\mathcal{S}&=-\tr[\rho\,\ln\,\rho] 
=-\sum_{n=0}^{\infty}\int_H {\rm d} \Pi_n\,\frac{\df P}{\df \Pi_n} \ln \Big(\frac{\df P}{\df \Pi_n}\Big)\,.
\end{align}  
This conforms to our expectation about the entropy of a decohered quantum system: it is simply the entropy of the quasi-classical probability distribution given by the diagonal matrix elements in the basis that diagonalizes the matrix.

{\it Leading logarithmic calculation. --}
Our goal will be to calculate the entropy at LL accuracy, where the LL subjet distribution is defined as 
\begin{align}\label{eq:log_expansion}
\frac{\df P}{\df \Pi_n}&=\frac{\df P_{\rm LL}}{\df \Pi_n}\Big(\alpha_s\ln\frac{R}{R_c}\ln\frac{E}{\Ecut}\Big)
\nn \\ & \quad \times
\Big[1+\mathcal{O}\Big(\alpha_s \ln \frac{R}{R_c}, \al_s \ln\frac{E}{\Ecut}\Big)\Big]
\,,\end{align}
with $E$ the energy and $R$ the radius parameter of the (fat) parent jet.
We use the angular-ordered factorization found in~\cite{Bassetto:1982ma,Bassetto:1984ik}: a hard parton, of flavor $i$ and energy $E$ produced in the short-distance scattering, undergoes a series of splittings, which have a much smaller angle than the previous one. Thus the initial parton $i$ splits at an angle $\theta$ into daughters with flavors $j$ and $k$ carrying a fraction $z$ and $1-z$ of the initial momentum, as described by
\begin{align}\label{eq:angular_ordering_factorization}
  \frac{\df P^{i}}{\df\Pi_n}(E,R)&= \frac{4\pi^2}{z(1-z) \theta E^2} \sum_{j,k}\frac{\alpha_s}{\pi}\frac{\mathcal{P}_{i\rightarrow jk}(z)}{2\theta} e^{-\Delta_i(R,\theta)}
  \nn \\ & \quad \times
  \frac{\df P^j}{\df\Pi_m}(zE,\theta)\,\frac{\df P^k}{\df\Pi_{n-m}}\bigl((1-z)E,\theta\bigr)
  \nn \,,\\
  \Delta_i(\Rmax,\theta)&=\sum_{j,k}\int_{\zcut}^{1-\zcut}\!\! \df z\int_{\theta}^{R}\!\df \theta'\,\frac{\alpha_s}{\pi}\,\frac{\mathcal{P}_{i\rightarrow jk}(z)}{2\theta'}
\,.\end{align}
Here $\mathcal{P}_{i\rightarrow jk}$ are splitting functions, $\Delta_i(\Rmax,\theta)$ is a Sudakov factor describing the no-splitting probability between the angle $R$ and $\theta$, $\zcut = \Ecut/E$, the indices $j,k$ denote all possible flavor combinations, and the strong coupling $\alpha_s$ is evaluated at the scale set by the transverse momentum of the splitting. The overall factor cancels a corresponding factor in the phase space in \eq{phase_space_factorization}, but enters in the entropy because of the logarithm in \eq{von_neumann_entropy}. We assumed that one daughter will split into $m$ partons and the other daughter into $n-m$ partons, and it is crucial that the sum on $m$ is part of the phase space so it sits in front (rather than inside) the logarithm in \eq{von_neumann_entropy}. The reason is that a parton produced by daughter $j$, has an angle with $j$ much smaller than $\theta$ due to strong angular ordering, and can therefore never be produced by parton $k$. Not all contributions to the cross section satisfy strong angular ordering, but these are subleading in the expansion of \eq{log_expansion}. 

Corresponding to the factorization of the probability densities, the phase space of the $n$-hard emissions factorizes as
\begin{align}\label{eq:phase_space_factorization}
\int_H d\Pi_n(p_J,R)&= \sum_{m=1}^{n-1}\int_{\zcut}^{1-\zcut}\!\df z\int_{\Rcut}^{\Rmax}\!\df\theta\,
\frac{z(1-z) \theta E^2}{4\pi^2}
\nn \\ & \quad \times
\int_H \df \Pi_m(zp_J+q_\perp,\theta)
\nn \\ & \quad \times
\int_H \df \Pi_{n-m}((1-z)p_J-q_\perp,\theta)\,.
\end{align}
Here we integrate over the momentum fractions and splitting angles for the splitting of the initial parton, sum over all partitions $m$ of the daughters, and integrate over their phase space. The upper bound of subsequent splittings is $\theta$ rather than the jet radius $R$, as indicated by the second argument of the phase space. The transverse momentum $q_\perp$ of the daughters is related to the angle $\theta$ by $|q_\perp| = z(1-z) E \theta$.

Combining Eqs.~\eqref{eq:von_neumann_entropy}, \eqref{eq:angular_ordering_factorization}, and \eqref{eq:phase_space_factorization}, and taking the soft limit $z \ll 1$ of the splitting functions, we get the following equation for the entropy
\begin{widetext}
    \begin{align}\label{eq:entropy}
      \mathcal{S}_i(E,R)&=\mathcal{F}_i(E,R)+ \int_{\zcut}^{1}\frac{\df z}{z}\int_{\Rcut}^{\Rmax}\frac{\df\theta}{\theta}\,\frac{2\alpha_s(zE\theta)C_i}{\pi}\,e^{-\Delta_{i}(\Rmax,\theta)}\big[\mathcal{S}_{g}(zE,\theta)+\mathcal{S}_{i}(E,\theta)\big]\,,
     \nn \\
      \mathcal{F}_i(E,R)&=\Delta_i(R,R_c)\,e^{-\Delta_{i}(\Rmax,\Rcut)}
      +   \int_{\zcut}^{1}\frac{\df z}{z}\int_{\Rcut}^{\Rmax}\frac{\df\theta}{\theta}\,\frac{2\alpha_s(zE\theta)C_i}{\pi}\,e^{-\Delta_{i}(\Rmax,\theta)}\Big[\Delta_{i}(\Rmax,\theta)-\ln\Big(\frac{8\pi C_i\alpha_s(zE\theta) \Lambda^2}{z^2\theta^2 E^2}\Big)\Big]
    \,.\end{align}
\end{widetext}
Here $C_q = \tfrac43$ for (anti-)quarks and $C_g = 3$ for gluons. We are forced to introduce the energy scale $\Lambda$, to make the probability entering the logarithm in \eq{von_neumann_entropy} dimensionless, converting the phase-space volume into a number of states. It is natural to take $\Lambda \propto E_c R_c$, using the phase-space volume of the infrared states that are traced over as unit. Note that in the soft limit the $g\rightarrow q\bar{q}$ splitting is subleading, and that we can replace $1-z \to 1$.

Focusing on a gluon jet, we first multiply both sides of \eq{entropy} by the inverse Sudakov factor $e^{\Delta_g(R,\Rcut)}$, and take the derivative with respect to $R$, to obtain
\begin{align}\label{eq:entropy_glue}
         R\frac{\partial \mathcal{S}_g}{\partial R}(E,R)&=e^{-\Delta_{g}(\Rmax,\Rcut)}\,R\frac{\partial}{\partial R}\Big(e^{\Delta_{g}(\Rmax,\Rcut)}\mathcal{F}_g(E,R)\Big)
 \nn\\ &\quad
+\int_{\zcut}^1\frac{\df z}{z}\,\frac{2\alpha_s(zE R)C_A}{\pi}\,\mathcal{S}_g(zE,R)\,.
\end{align}
We can obtain an analytical solution by ignoring the running of the coupling, in which case the first term in \eq{entropy_glue} simplifies significantly, leading to
\begin{align}\label{eq:entropy_glue_solution}
  \mathcal{S}_g(E,R)& = \Big(1+\ln \frac{E_c^2 R_c^2}{8\pi C_A \alpha_s \Lambda^2}\Big) \bigl(I_0(x)-1\bigr)
  \nn \\ & \quad
  + 2\ln \frac{ER}{E_c R_c} \Big(\frac{2}{x} I_1(x) - 1 \Big)
  \,,\nn \\ 
  x &\equiv 2 \Big[\frac{2 \al_s C_A}{\pi}\, \ln \frac{E}{E_c} \ln \frac{R}{R_c}\Big]^{1/2}
\,,\end{align}
where $I_0$ and $I_1$ are modified Bessel functions.

The evolution equations in \eq{entropy} resemble those obeyed by multiplicity of (sub-)jets or hadrons~\cite{Mueller:1982cq,Catani:1991pm,Khoze:2000iq,Gerwick:2012fw}. At LL accuracy the multiplicity is described by $I_0(x)$, and the first line of \eq{entropy_glue_solution} can be identified as a driving term proportional to the number of branchings (equal to the multiplicity minus 1). We find asymptotically the same growth in entropy as in the average subjet multiplicity:
\begin{align}
         \mathcal{S}_g(E,R)&\propto \frac{e^x}{\sqrt{x}}\,.
\end{align}

Under the conformal mapping that relates partons showers to initial-state cascades used in forward-scattering descriptions of deep-inelastic scattering~\cite{Hatta:2008st}, we can compare to the entropy of a density matrix resulting from tracing over color-connected dipoles of too small transverse size in impact-parameter space. Under this mapping, the energy ordering ($\ln({E}/{\Ecut})$) of the parton shower corresponds to the rapidity ordering ($Y$) of the initial-state cascade, while angles correspond to the transverse size of the dipole in impact-parameter space. Though the initial conditions are radically different,  based on the conformal mapping we conjecture that the entropy grows as $e^{\beta\sqrt{Y}}$ with $\beta$ a constant. Interestingly, this differs from \cite{Kharzeev:2017qzs}, where the entropy was estimated to grow proportional to $Y$, and \cite{Hagiwara:2017uaz}, finding an exponential growth $e^{\beta Y}$. However, it is not clear that the conformal mapping translates the decomposition of the Hilbert space used to define the reduced density matrix for jet physics into the same one used in forward-scattering. 
         
{\it Numerical results. --} 
The entropy can also be considered an example of a fractal jet observable. Those observables depend on the clustering tree of a jet algorithm~\cite{Elder:2017bkd}. For the entropy we use the Cambridge/Aachen algorithm~\cite{Ellis:1993tq}, which combines the two particles closest in angle into a parent pseudoparticle (by summing their momenta), repeating until the list consists of a single pseudoparticle. This can be thought of as treating the two nearest particles as arising from a perturbative splitting, which is in accord with the angular-ordered approximation we employed in the leading logarithmic calculation of the entropy, i.e.~the C/A clustering tree corresponds to the branching history. 

This branching history yields a list of energy fractions $Z=\{z_1,...,z_n\}$ (taking the smaller of the energy fractions of the daughters being clustered, defined with respect to the jet energy), and a list of branching angles $\Theta=\{\theta_1,...,\theta_n\}$ (the relative angles between the daughters clustered in each step). They are reversely ordered compared to the C/A clustering, so $\theta_1>\theta_2>...>\theta_n$. These lists correspond to the branchings which generate hard subjets, so that $\theta_n>\Rcut$, and $z_i E>\Ecut$ for each $z_i\in Z$. The multiplicity of hard subjets is therefore $n+1$, where the $+1$ comes from the initiating parton. At LL accuracy, we can compute the entropy as 
\begin{align}\label{eq:shower_entropy}
  s_g(Z,\Theta) & = \Delta_g(R,\Rcut) + \sum_{j=1}^{n}\Big[\Delta_g(\theta_j,\Rcut; z_j, z_c)
  \nn \\ & \quad
  -\ln\Big(\frac{8\pi C_A\alpha_s(z_jE\theta_j) \Lambda^2}{z^2_j \theta^2_j E^2}\Big)\Big]\,,
\end{align}
which follows from \eq{entropy}.
Here $\Delta_g$ is the Sudakov factor (for simplicity, we consider only gluons), and the extra arguments $z_j$ and $z_c$ indicate the range of the $z$ integral. The interpretation of \eq{shower_entropy} is that the entropy of the shower is the sum of the entropies generated at each step of the shower. If we average over many events, denoted by  $\langle\!\langle \cdot\rangle\!\rangle$, this converges to \eq{entropy},
\begin{align}\label{eq:event_averaging}
  \mathcal{S}_g=\langle\!\langle s_g\rangle\!\rangle\,.
\end{align} 

\begin{figure}
\includegraphics[width=0.45\textwidth]{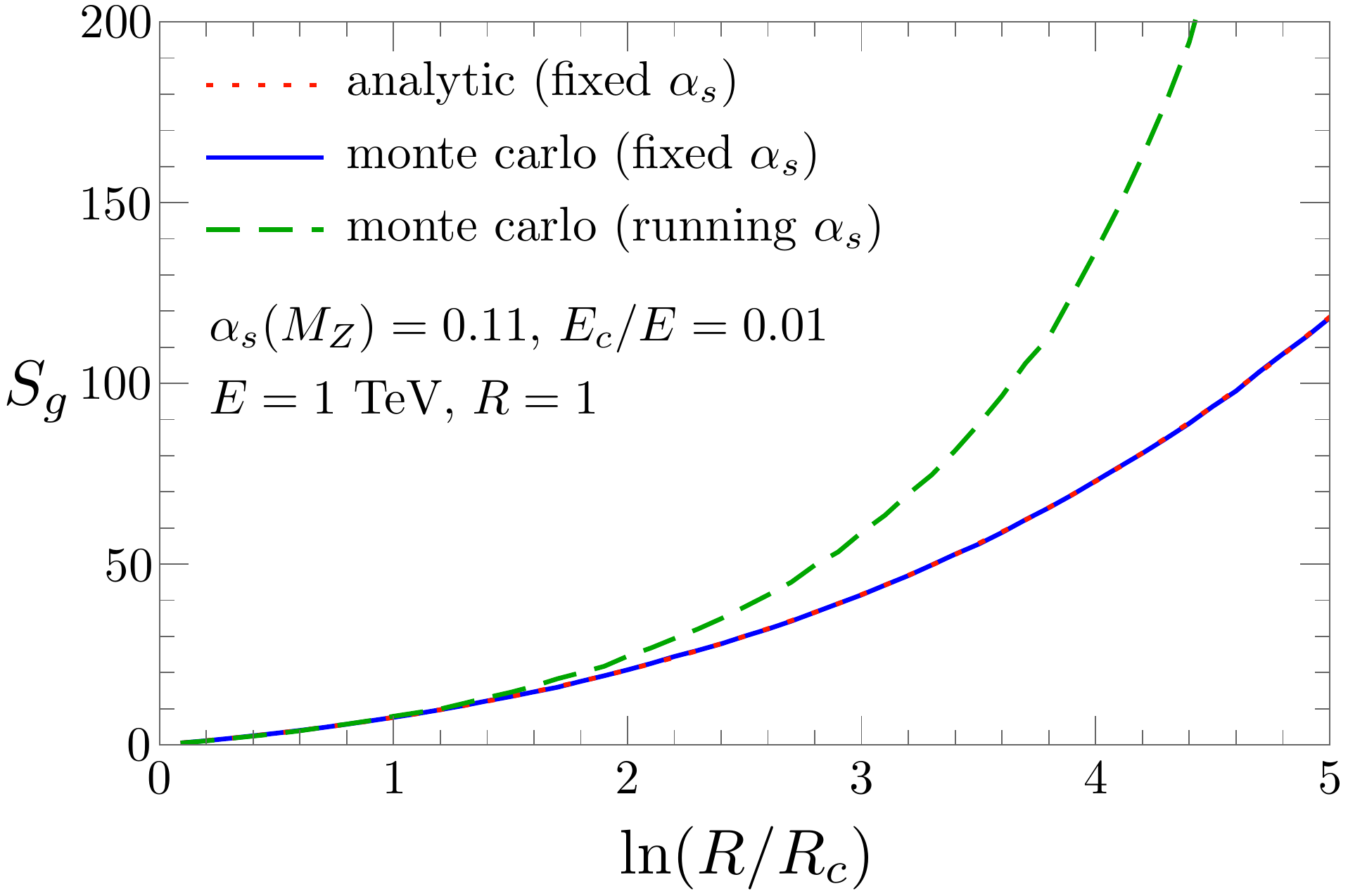}
\caption{The entropy of a gluon jet as function of the ratio of the jet radius $R$ over the subjet radius $R_c$, with $\Lambda^2=\frac{E_c^2 R_c^2}{8\pi C_A \alpha_s}$. Shown are the analytic (red dotted) and Monte Carlo (blue solid) result for $E/E_c=0.01$ at fixed coupling, as well as the Monte Carlo with running coupling (green dashed), which now depends on the jet energy $E=1$ TeV and radius $R=1$.}\label{fig:entropy}
\end{figure} 
In Fig.~\ref{fig:entropy} we show the entropy of a gluon jet in \eq{entropy_glue_solution} as the angular cutoff $\Rcut$ is lowered, comparing against a LL shower with both fixed and running coupling. This shower uses the collinear-soft approximation to the branching amplitudes, and ignores recoil effects. We take $\Lambda^2=E_c^2 R_c^2/(8\pi C_A \alpha_s)$, but other choices would simply add to our result a constant times the multiplicity minus one. At fixed coupling, we see exact agreement with our analytic expressions. Using the shower we can gauge the effect of the running coupling, with substantial deviations as the Landau pole is approached on the right side of the plot. With the appropriate generalization of the definitions in \eqs{shower_entropy}{event_averaging}, the entropy generated in an arbitrary parton shower or experiment can be measured. 

{\it Conclusions. --}
Fundamentally, the generation of the final state in a scattering process in an asymptotically-free gauge theory (or in QED or gravity) is a stochastic process, driven by the underlying quantum dynamics of the field theory, decohering the produced hard states, such that superpositions of momentum states are destroyed, quantum intereference effects are lost, and a quasi-classical probability distribution dominates. This is characterized by a density matrix with a non-zero entropy, which we calculated at LL accuracy for jets. In the approximations used for calculating scattering cross sections, the hard states are always fully decohered, regardless of the resolution parameter: at any finite resolution of the states, the measurement is formally integrated over an infinitely long time (see \cite{Sveshnikov:1995vi}), allowing the production of ever softer and more collinear radiation below the resolution scale.  In the LL picture, this stochastic process has a ``time'' associated to it, evolving from the widest angles that the hard partons can emit down to the smallest angles. We have shown that this evolution satisfies a second law: entropy increases as we examine the final state at smaller angles.

Further, the entropy obeys an evolution equation closely related to the subjet multiplicity, approaching it asymptotically (up to a constant). This is perhaps not surprising, given that contribution to the entropy at each splitting is determined by the available phase space. Thus the entropy of the process creating the jet should be connected to the so-called $\lambda$-measure introduced in \cite{Andersson:1988ee,Andersson:1989ww} as a proxy for the multiplicity, as well as a means to investigate the fractal nature of how the parton shower distributes the momentum of the initial hard state into smaller phase-space cells~\cite{Dahlqvist:1989yc,Gustafson:1991ru,Davighi:2017hok}.

The growth in entropy has a practical consequence for jet substructure and, in particular, machine learning. One can train discriminators based on course-grained representations, truncating the energy flow to only a few momentum regions (see, e.g.,~\cite{Datta:2017rhs,Dreyer:2018nbf,Chien:2018dfn}), a much larger basis \cite{Komiske:2017aww}, or as fine grained a representation as experimentally possible (the ``jet image''~\cite{Cogan:2014oua,deOliveira:2015xxd,Baldi:2016fql,Guest:2016iqz}). All approaches can provide similar discrimination power, even though it may seem that the latter contains more information. However, the same amount of information has just been further stochastically diluted, explaining why machine learning discriminators can saturate with only a relatively small number of momentum regions. There is of course a compensating effect, as coarse graining can deteriorate the angular and energy resolution. Additionally, there is something to be learned from the scaling pattern of the entropy as a fractal observable, and it would be fascinating to see how such an observable is related to machine learning discriminators.

D.N.~thanks Varun Vaidya, Stefan Prestel, Ira Rothstein and Dan Carney for helpful discussions. We thank Jesse Thaler for feedback on this Letter. We acknowledge support from DOE contract DE-AC52-06NA25396, the LANL/LDRD Program, the ERC Grant ERC-STG-2015-677323 and the D-ITP consortium, a program of the Netherlands Organization for Scientific Research (NWO) that is funded by the Dutch Ministry of Education, Culture and Science (OCW). W.W.~thanks the LANL theory group for hospitality during the initial phase of this project.

\bibliography{Entropy_of_Jets.bib}

\begin{thebibliography}{71}%
\makeatletter
\providecommand \@ifxundefined [1]{%
 \@ifx{#1\undefined}
}%
\providecommand \@ifnum [1]{%
 \ifnum #1\expandafter \@firstoftwo
 \else \expandafter \@secondoftwo
 \fi
}%
\providecommand \@ifx [1]{%
 \ifx #1\expandafter \@firstoftwo
 \else \expandafter \@secondoftwo
 \fi
}%
\providecommand \natexlab [1]{#1}%
\providecommand \enquote  [1]{``#1''}%
\providecommand \bibnamefont  [1]{#1}%
\providecommand \bibfnamefont [1]{#1}%
\providecommand \citenamefont [1]{#1}%
\providecommand \href@noop [0]{\@secondoftwo}%
\providecommand \href [0]{\begingroup \@sanitize@url \@href}%
\providecommand \@href[1]{\@@startlink{#1}\@@href}%
\providecommand \@@href[1]{\endgroup#1\@@endlink}%
\providecommand \@sanitize@url [0]{\catcode `\\12\catcode `\$12\catcode
  `\&12\catcode `\#12\catcode `\^12\catcode `\_12\catcode `\%12\relax}%
\providecommand \@@startlink[1]{}%
\providecommand \@@endlink[0]{}%
\providecommand \url  [0]{\begingroup\@sanitize@url \@url }%
\providecommand \@url [1]{\endgroup\@href {#1}{\urlprefix }}%
\providecommand \urlprefix  [0]{URL }%
\providecommand \Eprint [0]{\href }%
\providecommand \doibase [0]{http://dx.doi.org/}%
\providecommand \selectlanguage [0]{\@gobble}%
\providecommand \bibinfo  [0]{\@secondoftwo}%
\providecommand \bibfield  [0]{\@secondoftwo}%
\providecommand \translation [1]{[#1]}%
\providecommand \BibitemOpen [0]{}%
\providecommand \bibitemStop [0]{}%
\providecommand \bibitemNoStop [0]{.\EOS\space}%
\providecommand \EOS [0]{\spacefactor3000\relax}%
\providecommand \BibitemShut  [1]{\csname bibitem#1\endcsname}%
\let\auto@bib@innerbib\@empty
\bibitem [{\citenamefont {Hawking}\ \emph {et~al.}(2016)\citenamefont
  {Hawking}, \citenamefont {Perry},\ and\ \citenamefont
  {Strominger}}]{Hawking:2016msc}%
  \BibitemOpen
  \bibfield  {author} {\bibinfo {author} {\bibfnamefont {S.~W.}\ \bibnamefont
  {Hawking}}, \bibinfo {author} {\bibfnamefont {M.~J.}\ \bibnamefont {Perry}},
  \ and\ \bibinfo {author} {\bibfnamefont {A.}~\bibnamefont {Strominger}},\
  }\href {\doibase 10.1103/PhysRevLett.116.231301} {\bibfield  {journal}
  {\bibinfo  {journal} {Phys. Rev. Lett.}\ }\textbf {\bibinfo {volume} {116}},\
  \bibinfo {pages} {231301} (\bibinfo {year} {2016})},\ \Eprint
  {http://arxiv.org/abs/1601.00921} {arXiv:1601.00921 [hep-th]} \BibitemShut
  {NoStop}%
\bibitem [{\citenamefont {Haco}\ \emph {et~al.}(2018)\citenamefont {Haco},
  \citenamefont {Hawking}, \citenamefont {Perry},\ and\ \citenamefont
  {Strominger}}]{Haco:2018ske}%
  \BibitemOpen
  \bibfield  {author} {\bibinfo {author} {\bibfnamefont {S.}~\bibnamefont
  {Haco}}, \bibinfo {author} {\bibfnamefont {S.~W.}\ \bibnamefont {Hawking}},
  \bibinfo {author} {\bibfnamefont {M.~J.}\ \bibnamefont {Perry}}, \ and\
  \bibinfo {author} {\bibfnamefont {A.}~\bibnamefont {Strominger}},\
  }\href@noop {} {\  (\bibinfo {year} {2018})},\ \Eprint
  {http://arxiv.org/abs/1810.01847} {arXiv:1810.01847 [hep-th]} \BibitemShut
  {NoStop}%
\bibitem [{\citenamefont {Berges}\ \emph {et~al.}(2018)\citenamefont {Berges},
  \citenamefont {Floerchinger},\ and\ \citenamefont
  {Venugopalan}}]{Berges:2017hne}%
  \BibitemOpen
  \bibfield  {author} {\bibinfo {author} {\bibfnamefont {J.}~\bibnamefont
  {Berges}}, \bibinfo {author} {\bibfnamefont {S.}~\bibnamefont
  {Floerchinger}}, \ and\ \bibinfo {author} {\bibfnamefont {R.}~\bibnamefont
  {Venugopalan}},\ }\href {\doibase 10.1007/JHEP04(2018)145} {\bibfield
  {journal} {\bibinfo  {journal} {JHEP}\ }\textbf {\bibinfo {volume} {04}},\
  \bibinfo {pages} {145} (\bibinfo {year} {2018})},\ \Eprint
  {http://arxiv.org/abs/1712.09362} {arXiv:1712.09362 [hep-th]} \BibitemShut
  {NoStop}%
\bibitem [{\citenamefont {Andersson}\ \emph {et~al.}(1983)\citenamefont
  {Andersson}, \citenamefont {Gustafson}, \citenamefont {Ingelman},\ and\
  \citenamefont {Sjostrand}}]{Andersson:1983ia}%
  \BibitemOpen
  \bibfield  {author} {\bibinfo {author} {\bibfnamefont {B.}~\bibnamefont
  {Andersson}}, \bibinfo {author} {\bibfnamefont {G.}~\bibnamefont
  {Gustafson}}, \bibinfo {author} {\bibfnamefont {G.}~\bibnamefont {Ingelman}},
  \ and\ \bibinfo {author} {\bibfnamefont {T.}~\bibnamefont {Sjostrand}},\
  }\href {\doibase 10.1016/0370-1573(83)90080-7} {\bibfield  {journal}
  {\bibinfo  {journal} {Phys. Rept.}\ }\textbf {\bibinfo {volume} {97}},\
  \bibinfo {pages} {31} (\bibinfo {year} {1983})}\BibitemShut {NoStop}%
\bibitem [{\citenamefont {Andersson}(1997)}]{Andersson:1998tv}%
  \BibitemOpen
  \bibfield  {author} {\bibinfo {author} {\bibfnamefont {B.}~\bibnamefont
  {Andersson}},\ }\href@noop {} {\bibfield  {journal} {\bibinfo  {journal}
  {Camb. Monogr. Part. Phys. Nucl. Phys. Cosmol.}\ }\textbf {\bibinfo {volume}
  {7}},\ \bibinfo {pages} {1} (\bibinfo {year} {1997})}\BibitemShut {NoStop}%
\bibitem [{\citenamefont {Larkoski}\ \emph {et~al.}(2017)\citenamefont
  {Larkoski}, \citenamefont {Moult},\ and\ \citenamefont
  {Nachman}}]{Larkoski:2017jix}%
  \BibitemOpen
  \bibfield  {author} {\bibinfo {author} {\bibfnamefont {A.~J.}\ \bibnamefont
  {Larkoski}}, \bibinfo {author} {\bibfnamefont {I.}~\bibnamefont {Moult}}, \
  and\ \bibinfo {author} {\bibfnamefont {B.}~\bibnamefont {Nachman}},\
  }\href@noop {} {\  (\bibinfo {year} {2017})},\ \Eprint
  {http://arxiv.org/abs/1709.04464} {arXiv:1709.04464 [hep-ph]} \BibitemShut
  {NoStop}%
\bibitem [{\citenamefont {Larkoski}\ \emph {et~al.}(2014)\citenamefont
  {Larkoski}, \citenamefont {Thaler},\ and\ \citenamefont
  {Waalewijn}}]{Larkoski:2014pca}%
  \BibitemOpen
  \bibfield  {author} {\bibinfo {author} {\bibfnamefont {A.~J.}\ \bibnamefont
  {Larkoski}}, \bibinfo {author} {\bibfnamefont {J.}~\bibnamefont {Thaler}}, \
  and\ \bibinfo {author} {\bibfnamefont {W.~J.}\ \bibnamefont {Waalewijn}},\
  }\href {\doibase 10.1007/JHEP11(2014)129} {\bibfield  {journal} {\bibinfo
  {journal} {JHEP}\ }\textbf {\bibinfo {volume} {11}},\ \bibinfo {pages} {129}
  (\bibinfo {year} {2014})},\ \Eprint {http://arxiv.org/abs/1408.3122}
  {arXiv:1408.3122 [hep-ph]} \BibitemShut {NoStop}%
\bibitem [{\citenamefont {Datta}\ and\ \citenamefont
  {Larkoski}(2017)}]{Datta:2017rhs}%
  \BibitemOpen
  \bibfield  {author} {\bibinfo {author} {\bibfnamefont {K.}~\bibnamefont
  {Datta}}\ and\ \bibinfo {author} {\bibfnamefont {A.}~\bibnamefont
  {Larkoski}},\ }\href {\doibase 10.1007/JHEP06(2017)073} {\bibfield  {journal}
  {\bibinfo  {journal} {JHEP}\ }\textbf {\bibinfo {volume} {06}},\ \bibinfo
  {pages} {073} (\bibinfo {year} {2017})},\ \Eprint
  {http://arxiv.org/abs/1704.08249} {arXiv:1704.08249 [hep-ph]} \BibitemShut
  {NoStop}%
\bibitem [{\citenamefont {Dasgupta}\ and\ \citenamefont
  {Salam}(2001)}]{Dasgupta:2001sh}%
  \BibitemOpen
  \bibfield  {author} {\bibinfo {author} {\bibfnamefont {M.}~\bibnamefont
  {Dasgupta}}\ and\ \bibinfo {author} {\bibfnamefont {G.~P.}\ \bibnamefont
  {Salam}},\ }\href {\doibase 10.1016/S0370-2693(01)00725-0} {\bibfield
  {journal} {\bibinfo  {journal} {Phys. Lett.}\ }\textbf {\bibinfo {volume}
  {B512}},\ \bibinfo {pages} {323} (\bibinfo {year} {2001})},\ \Eprint
  {http://arxiv.org/abs/hep-ph/0104277} {arXiv:hep-ph/0104277} \BibitemShut
  {NoStop}%
\bibitem [{\citenamefont {Banfi}\ \emph {et~al.}(2002)\citenamefont {Banfi},
  \citenamefont {Marchesini},\ and\ \citenamefont {Smye}}]{Banfi:2002hw}%
  \BibitemOpen
  \bibfield  {author} {\bibinfo {author} {\bibfnamefont {A.}~\bibnamefont
  {Banfi}}, \bibinfo {author} {\bibfnamefont {G.}~\bibnamefont {Marchesini}}, \
  and\ \bibinfo {author} {\bibfnamefont {G.}~\bibnamefont {Smye}},\ }\href
  {\doibase 10.1088/1126-6708/2002/08/006} {\bibfield  {journal} {\bibinfo
  {journal} {JHEP}\ }\textbf {\bibinfo {volume} {08}},\ \bibinfo {pages} {006}
  (\bibinfo {year} {2002})},\ \Eprint {http://arxiv.org/abs/hep-ph/0206076}
  {arXiv:hep-ph/0206076} \BibitemShut {NoStop}%
\bibitem [{\citenamefont {Weigert}(2004)}]{Weigert:2003mm}%
  \BibitemOpen
  \bibfield  {author} {\bibinfo {author} {\bibfnamefont {H.}~\bibnamefont
  {Weigert}},\ }\href {\doibase 10.1016/j.nuclphysb.2004.03.002} {\bibfield
  {journal} {\bibinfo  {journal} {Nucl. Phys.}\ }\textbf {\bibinfo {volume}
  {B685}},\ \bibinfo {pages} {321} (\bibinfo {year} {2004})},\ \Eprint
  {http://arxiv.org/abs/hep-ph/0312050} {arXiv:hep-ph/0312050} \BibitemShut
  {NoStop}%
\bibitem [{\citenamefont {McLerran}\ and\ \citenamefont
  {Venugopalan}(1994{\natexlab{a}})}]{McLerran:1993ka}%
  \BibitemOpen
  \bibfield  {author} {\bibinfo {author} {\bibfnamefont {L.~D.}\ \bibnamefont
  {McLerran}}\ and\ \bibinfo {author} {\bibfnamefont {R.}~\bibnamefont
  {Venugopalan}},\ }\href {\doibase 10.1103/PhysRevD.49.3352} {\bibfield
  {journal} {\bibinfo  {journal} {Phys. Rev.}\ }\textbf {\bibinfo {volume}
  {D49}},\ \bibinfo {pages} {3352} (\bibinfo {year} {1994}{\natexlab{a}})},\
  \Eprint {http://arxiv.org/abs/hep-ph/9311205} {arXiv:hep-ph/9311205}
  \BibitemShut {NoStop}%
\bibitem [{\citenamefont {McLerran}\ and\ \citenamefont
  {Venugopalan}(1994{\natexlab{b}})}]{McLerran:1993ni}%
  \BibitemOpen
  \bibfield  {author} {\bibinfo {author} {\bibfnamefont {L.~D.}\ \bibnamefont
  {McLerran}}\ and\ \bibinfo {author} {\bibfnamefont {R.}~\bibnamefont
  {Venugopalan}},\ }\href {\doibase 10.1103/PhysRevD.49.2233} {\bibfield
  {journal} {\bibinfo  {journal} {Phys. Rev.}\ }\textbf {\bibinfo {volume}
  {D49}},\ \bibinfo {pages} {2233} (\bibinfo {year} {1994}{\natexlab{b}})},\
  \Eprint {http://arxiv.org/abs/hep-ph/9309289} {arXiv:hep-ph/9309289}
  \BibitemShut {NoStop}%
\bibitem [{\citenamefont {McLerran}\ and\ \citenamefont
  {Venugopalan}(1994{\natexlab{c}})}]{McLerran:1994vd}%
  \BibitemOpen
  \bibfield  {author} {\bibinfo {author} {\bibfnamefont {L.~D.}\ \bibnamefont
  {McLerran}}\ and\ \bibinfo {author} {\bibfnamefont {R.}~\bibnamefont
  {Venugopalan}},\ }\href {\doibase 10.1103/PhysRevD.50.2225} {\bibfield
  {journal} {\bibinfo  {journal} {Phys. Rev.}\ }\textbf {\bibinfo {volume}
  {D50}},\ \bibinfo {pages} {2225} (\bibinfo {year} {1994}{\natexlab{c}})},\
  \Eprint {http://arxiv.org/abs/hep-ph/9402335} {arXiv:hep-ph/9402335}
  \BibitemShut {NoStop}%
\bibitem [{\citenamefont {Balitsky}(1996)}]{Balitsky:1995ub}%
  \BibitemOpen
  \bibfield  {author} {\bibinfo {author} {\bibfnamefont {I.}~\bibnamefont
  {Balitsky}},\ }\href {\doibase 10.1016/0550-3213(95)00638-9} {\bibfield
  {journal} {\bibinfo  {journal} {Nucl. Phys.}\ }\textbf {\bibinfo {volume}
  {B463}},\ \bibinfo {pages} {99} (\bibinfo {year} {1996})},\ \Eprint
  {http://arxiv.org/abs/hep-ph/9509348} {arXiv:hep-ph/9509348} \BibitemShut
  {NoStop}%
\bibitem [{\citenamefont {Kovchegov}(1999)}]{Kovchegov:1999yj}%
  \BibitemOpen
  \bibfield  {author} {\bibinfo {author} {\bibfnamefont {Y.~V.}\ \bibnamefont
  {Kovchegov}},\ }\href {\doibase 10.1103/PhysRevD.60.034008} {\bibfield
  {journal} {\bibinfo  {journal} {Phys. Rev.}\ }\textbf {\bibinfo {volume}
  {D60}},\ \bibinfo {pages} {034008} (\bibinfo {year} {1999})},\ \Eprint
  {http://arxiv.org/abs/hep-ph/9901281} {arXiv:hep-ph/9901281} \BibitemShut
  {NoStop}%
\bibitem [{\citenamefont {Jalilian-Marian}\ \emph {et~al.}(1997)\citenamefont
  {Jalilian-Marian}, \citenamefont {Kovner}, \citenamefont {McLerran},\ and\
  \citenamefont {Weigert}}]{JalilianMarian:1996xn}%
  \BibitemOpen
  \bibfield  {author} {\bibinfo {author} {\bibfnamefont {J.}~\bibnamefont
  {Jalilian-Marian}}, \bibinfo {author} {\bibfnamefont {A.}~\bibnamefont
  {Kovner}}, \bibinfo {author} {\bibfnamefont {L.~D.}\ \bibnamefont
  {McLerran}}, \ and\ \bibinfo {author} {\bibfnamefont {H.}~\bibnamefont
  {Weigert}},\ }\href {\doibase 10.1103/PhysRevD.55.5414} {\bibfield  {journal}
  {\bibinfo  {journal} {Phys. Rev.}\ }\textbf {\bibinfo {volume} {D55}},\
  \bibinfo {pages} {5414} (\bibinfo {year} {1997})},\ \Eprint
  {http://arxiv.org/abs/hep-ph/9606337} {arXiv:hep-ph/9606337} \BibitemShut
  {NoStop}%
\bibitem [{\citenamefont {Jalilian-Marian}\ \emph {et~al.}(1998)\citenamefont
  {Jalilian-Marian}, \citenamefont {Kovner}, \citenamefont {Leonidov},\ and\
  \citenamefont {Weigert}}]{JalilianMarian:1997gr}%
  \BibitemOpen
  \bibfield  {author} {\bibinfo {author} {\bibfnamefont {J.}~\bibnamefont
  {Jalilian-Marian}}, \bibinfo {author} {\bibfnamefont {A.}~\bibnamefont
  {Kovner}}, \bibinfo {author} {\bibfnamefont {A.}~\bibnamefont {Leonidov}}, \
  and\ \bibinfo {author} {\bibfnamefont {H.}~\bibnamefont {Weigert}},\ }\href
  {\doibase 10.1103/PhysRevD.59.014014} {\bibfield  {journal} {\bibinfo
  {journal} {Phys. Rev.}\ }\textbf {\bibinfo {volume} {D59}},\ \bibinfo {pages}
  {014014} (\bibinfo {year} {1998})},\ \Eprint
  {http://arxiv.org/abs/hep-ph/9706377} {arXiv:hep-ph/9706377} \BibitemShut
  {NoStop}%
\bibitem [{\citenamefont {Iancu}\ \emph {et~al.}(2001)\citenamefont {Iancu},
  \citenamefont {Leonidov},\ and\ \citenamefont {McLerran}}]{Iancu:2001ad}%
  \BibitemOpen
  \bibfield  {author} {\bibinfo {author} {\bibfnamefont {E.}~\bibnamefont
  {Iancu}}, \bibinfo {author} {\bibfnamefont {A.}~\bibnamefont {Leonidov}}, \
  and\ \bibinfo {author} {\bibfnamefont {L.~D.}\ \bibnamefont {McLerran}},\
  }\href {\doibase 10.1016/S0370-2693(01)00524-X} {\bibfield  {journal}
  {\bibinfo  {journal} {Phys. Lett.}\ }\textbf {\bibinfo {volume} {B510}},\
  \bibinfo {pages} {133} (\bibinfo {year} {2001})},\ \Eprint
  {http://arxiv.org/abs/hep-ph/0102009} {arXiv:hep-ph/0102009} \BibitemShut
  {NoStop}%
\bibitem [{\citenamefont {Hatta}(2008)}]{Hatta:2008st}%
  \BibitemOpen
  \bibfield  {author} {\bibinfo {author} {\bibfnamefont {Y.}~\bibnamefont
  {Hatta}},\ }\href {\doibase 10.1088/1126-6708/2008/11/057} {\bibfield
  {journal} {\bibinfo  {journal} {JHEP}\ }\textbf {\bibinfo {volume} {11}},\
  \bibinfo {pages} {057} (\bibinfo {year} {2008})},\ \Eprint
  {http://arxiv.org/abs/0810.0889} {arXiv:0810.0889 [hep-ph]} \BibitemShut
  {NoStop}%
\bibitem [{\citenamefont {Avsar}\ \emph {et~al.}(2009)\citenamefont {Avsar},
  \citenamefont {Hatta},\ and\ \citenamefont {Matsuo}}]{Avsar:2009yb}%
  \BibitemOpen
  \bibfield  {author} {\bibinfo {author} {\bibfnamefont {E.}~\bibnamefont
  {Avsar}}, \bibinfo {author} {\bibfnamefont {Y.}~\bibnamefont {Hatta}}, \ and\
  \bibinfo {author} {\bibfnamefont {T.}~\bibnamefont {Matsuo}},\ }\href
  {\doibase 10.1088/1126-6708/2009/06/011} {\bibfield  {journal} {\bibinfo
  {journal} {JHEP}\ }\textbf {\bibinfo {volume} {06}},\ \bibinfo {pages} {011}
  (\bibinfo {year} {2009})},\ \Eprint {http://arxiv.org/abs/0903.4285}
  {arXiv:0903.4285 [hep-ph]} \BibitemShut {NoStop}%
\bibitem [{\citenamefont {Caron-Huot}(2018)}]{Caron-Huot:2015bja}%
  \BibitemOpen
  \bibfield  {author} {\bibinfo {author} {\bibfnamefont {S.}~\bibnamefont
  {Caron-Huot}},\ }\href {\doibase 10.1007/JHEP03(2018)036} {\bibfield
  {journal} {\bibinfo  {journal} {JHEP}\ }\textbf {\bibinfo {volume} {03}},\
  \bibinfo {pages} {036} (\bibinfo {year} {2018})},\ \Eprint
  {http://arxiv.org/abs/1501.03754} {arXiv:1501.03754 [hep-ph]} \BibitemShut
  {NoStop}%
\bibitem [{\citenamefont {Kutak}(2011)}]{Kutak:2011rb}%
  \BibitemOpen
  \bibfield  {author} {\bibinfo {author} {\bibfnamefont {K.}~\bibnamefont
  {Kutak}},\ }\href {\doibase 10.1016/j.physletb.2011.09.113} {\bibfield
  {journal} {\bibinfo  {journal} {Phys. Lett.}\ }\textbf {\bibinfo {volume}
  {B705}},\ \bibinfo {pages} {217} (\bibinfo {year} {2011})},\ \Eprint
  {http://arxiv.org/abs/1103.3654} {arXiv:1103.3654 [hep-ph]} \BibitemShut
  {NoStop}%
\bibitem [{\citenamefont {Kovner}\ and\ \citenamefont
  {Lublinsky}(2015)}]{Kovner:2015hga}%
  \BibitemOpen
  \bibfield  {author} {\bibinfo {author} {\bibfnamefont {A.}~\bibnamefont
  {Kovner}}\ and\ \bibinfo {author} {\bibfnamefont {M.}~\bibnamefont
  {Lublinsky}},\ }\href {\doibase 10.1103/PhysRevD.92.034016} {\bibfield
  {journal} {\bibinfo  {journal} {Phys. Rev.}\ }\textbf {\bibinfo {volume}
  {D92}},\ \bibinfo {pages} {034016} (\bibinfo {year} {2015})},\ \Eprint
  {http://arxiv.org/abs/1506.05394} {arXiv:1506.05394 [hep-ph]} \BibitemShut
  {NoStop}%
\bibitem [{\citenamefont {Kharzeev}\ and\ \citenamefont
  {Levin}(2017)}]{Kharzeev:2017qzs}%
  \BibitemOpen
  \bibfield  {author} {\bibinfo {author} {\bibfnamefont {D.~E.}\ \bibnamefont
  {Kharzeev}}\ and\ \bibinfo {author} {\bibfnamefont {E.~M.}\ \bibnamefont
  {Levin}},\ }\href {\doibase 10.1103/PhysRevD.95.114008} {\bibfield  {journal}
  {\bibinfo  {journal} {Phys. Rev.}\ }\textbf {\bibinfo {volume} {D95}},\
  \bibinfo {pages} {114008} (\bibinfo {year} {2017})},\ \Eprint
  {http://arxiv.org/abs/1702.03489} {arXiv:1702.03489 [hep-ph]} \BibitemShut
  {NoStop}%
\bibitem [{\citenamefont {Hagiwara}\ \emph {et~al.}(2018)\citenamefont
  {Hagiwara}, \citenamefont {Hatta}, \citenamefont {Xiao},\ and\ \citenamefont
  {Yuan}}]{Hagiwara:2017uaz}%
  \BibitemOpen
  \bibfield  {author} {\bibinfo {author} {\bibfnamefont {Y.}~\bibnamefont
  {Hagiwara}}, \bibinfo {author} {\bibfnamefont {Y.}~\bibnamefont {Hatta}},
  \bibinfo {author} {\bibfnamefont {B.-W.}\ \bibnamefont {Xiao}}, \ and\
  \bibinfo {author} {\bibfnamefont {F.}~\bibnamefont {Yuan}},\ }\href {\doibase
  10.1103/PhysRevD.97.094029} {\bibfield  {journal} {\bibinfo  {journal} {Phys.
  Rev.}\ }\textbf {\bibinfo {volume} {D97}},\ \bibinfo {pages} {094029}
  (\bibinfo {year} {2018})},\ \Eprint {http://arxiv.org/abs/1801.00087}
  {arXiv:1801.00087 [hep-ph]} \BibitemShut {NoStop}%
\bibitem [{\citenamefont {Kovner}\ \emph {et~al.}(2018)\citenamefont {Kovner},
  \citenamefont {Lublinsky},\ and\ \citenamefont {Serino}}]{Kovner:2018rbf}%
  \BibitemOpen
  \bibfield  {author} {\bibinfo {author} {\bibfnamefont {A.}~\bibnamefont
  {Kovner}}, \bibinfo {author} {\bibfnamefont {M.}~\bibnamefont {Lublinsky}}, \
  and\ \bibinfo {author} {\bibfnamefont {M.}~\bibnamefont {Serino}},\
  }\href@noop {} {\  (\bibinfo {year} {2018})},\ \Eprint
  {http://arxiv.org/abs/1806.01089} {arXiv:1806.01089 [hep-ph]} \BibitemShut
  {NoStop}%
\bibitem [{\citenamefont {Balasubramanian}\ \emph {et~al.}(2012)\citenamefont
  {Balasubramanian}, \citenamefont {McDermott},\ and\ \citenamefont
  {Van~Raamsdonk}}]{Balasubramanian:2011wt}%
  \BibitemOpen
  \bibfield  {author} {\bibinfo {author} {\bibfnamefont {V.}~\bibnamefont
  {Balasubramanian}}, \bibinfo {author} {\bibfnamefont {M.~B.}\ \bibnamefont
  {McDermott}}, \ and\ \bibinfo {author} {\bibfnamefont {M.}~\bibnamefont
  {Van~Raamsdonk}},\ }\href {\doibase 10.1103/PhysRevD.86.045014} {\bibfield
  {journal} {\bibinfo  {journal} {Phys. Rev.}\ }\textbf {\bibinfo {volume}
  {D86}},\ \bibinfo {pages} {045014} (\bibinfo {year} {2012})},\ \Eprint
  {http://arxiv.org/abs/1108.3568} {arXiv:1108.3568 [hep-th]} \BibitemShut
  {NoStop}%
\bibitem [{\citenamefont {Hsu}\ \emph {et~al.}(2013)\citenamefont {Hsu},
  \citenamefont {McDermott},\ and\ \citenamefont {Van~Raamsdonk}}]{Hsu:2012gk}%
  \BibitemOpen
  \bibfield  {author} {\bibinfo {author} {\bibfnamefont {T.-C.~L.}\
  \bibnamefont {Hsu}}, \bibinfo {author} {\bibfnamefont {M.~B.}\ \bibnamefont
  {McDermott}}, \ and\ \bibinfo {author} {\bibfnamefont {M.}~\bibnamefont
  {Van~Raamsdonk}},\ }\href {\doibase 10.1007/JHEP11(2013)121} {\bibfield
  {journal} {\bibinfo  {journal} {JHEP}\ }\textbf {\bibinfo {volume} {11}},\
  \bibinfo {pages} {121} (\bibinfo {year} {2013})},\ \Eprint
  {http://arxiv.org/abs/1210.0054} {arXiv:1210.0054 [hep-th]} \BibitemShut
  {NoStop}%
\bibitem [{\citenamefont {Lello}\ \emph {et~al.}(2013)\citenamefont {Lello},
  \citenamefont {Boyanovsky},\ and\ \citenamefont {Holman}}]{Lello:2013bva}%
  \BibitemOpen
  \bibfield  {author} {\bibinfo {author} {\bibfnamefont {L.}~\bibnamefont
  {Lello}}, \bibinfo {author} {\bibfnamefont {D.}~\bibnamefont {Boyanovsky}}, \
  and\ \bibinfo {author} {\bibfnamefont {R.}~\bibnamefont {Holman}},\ }\href
  {\doibase 10.1007/JHEP11(2013)116} {\bibfield  {journal} {\bibinfo  {journal}
  {JHEP}\ }\textbf {\bibinfo {volume} {11}},\ \bibinfo {pages} {116} (\bibinfo
  {year} {2013})},\ \Eprint {http://arxiv.org/abs/1304.6110} {arXiv:1304.6110
  [hep-th]} \BibitemShut {NoStop}%
\bibitem [{\citenamefont {Nagy}\ and\ \citenamefont
  {Soper}(2007)}]{Nagy:2007ty}%
  \BibitemOpen
  \bibfield  {author} {\bibinfo {author} {\bibfnamefont {Z.}~\bibnamefont
  {Nagy}}\ and\ \bibinfo {author} {\bibfnamefont {D.~E.}\ \bibnamefont
  {Soper}},\ }\href {\doibase 10.1088/1126-6708/2007/09/114} {\bibfield
  {journal} {\bibinfo  {journal} {JHEP}\ }\textbf {\bibinfo {volume} {09}},\
  \bibinfo {pages} {114} (\bibinfo {year} {2007})},\ \Eprint
  {http://arxiv.org/abs/0706.0017} {arXiv:0706.0017 [hep-ph]} \BibitemShut
  {NoStop}%
\bibitem [{\citenamefont {Nagy}\ and\ \citenamefont
  {Soper}(2014)}]{Nagy:2014mqa}%
  \BibitemOpen
  \bibfield  {author} {\bibinfo {author} {\bibfnamefont {Z.}~\bibnamefont
  {Nagy}}\ and\ \bibinfo {author} {\bibfnamefont {D.~E.}\ \bibnamefont
  {Soper}},\ }\href {\doibase 10.1007/JHEP06(2014)097} {\bibfield  {journal}
  {\bibinfo  {journal} {JHEP}\ }\textbf {\bibinfo {volume} {06}},\ \bibinfo
  {pages} {097} (\bibinfo {year} {2014})},\ \Eprint
  {http://arxiv.org/abs/1401.6364} {arXiv:1401.6364 [hep-ph]} \BibitemShut
  {NoStop}%
\bibitem [{\citenamefont {Ellis}\ and\ \citenamefont
  {Soper}(1993)}]{Ellis:1993tq}%
  \BibitemOpen
  \bibfield  {author} {\bibinfo {author} {\bibfnamefont {S.~D.}\ \bibnamefont
  {Ellis}}\ and\ \bibinfo {author} {\bibfnamefont {D.~E.}\ \bibnamefont
  {Soper}},\ }\href {\doibase 10.1103/PhysRevD.48.3160} {\bibfield  {journal}
  {\bibinfo  {journal} {Phys. Rev.}\ }\textbf {\bibinfo {volume} {D48}},\
  \bibinfo {pages} {3160} (\bibinfo {year} {1993})},\ \Eprint
  {http://arxiv.org/abs/hep-ph/9305266} {arXiv:hep-ph/9305266} \BibitemShut
  {NoStop}%
\bibitem [{\citenamefont {Dokshitzer}\ \emph {et~al.}(1997)\citenamefont
  {Dokshitzer}, \citenamefont {Leder}, \citenamefont {Moretti},\ and\
  \citenamefont {Webber}}]{Dokshitzer:1997in}%
  \BibitemOpen
  \bibfield  {author} {\bibinfo {author} {\bibfnamefont {Y.~L.}\ \bibnamefont
  {Dokshitzer}}, \bibinfo {author} {\bibfnamefont {G.~D.}\ \bibnamefont
  {Leder}}, \bibinfo {author} {\bibfnamefont {S.}~\bibnamefont {Moretti}}, \
  and\ \bibinfo {author} {\bibfnamefont {B.~R.}\ \bibnamefont {Webber}},\
  }\href {\doibase 10.1088/1126-6708/1997/08/001} {\bibfield  {journal}
  {\bibinfo  {journal} {JHEP}\ }\textbf {\bibinfo {volume} {08}},\ \bibinfo
  {pages} {001} (\bibinfo {year} {1997})},\ \Eprint
  {http://arxiv.org/abs/hep-ph/9707323} {arXiv:hep-ph/9707323} \BibitemShut
  {NoStop}%
\bibitem [{\citenamefont {Sveshnikov}\ and\ \citenamefont
  {Tkachov}(1996)}]{Sveshnikov:1995vi}%
  \BibitemOpen
  \bibfield  {author} {\bibinfo {author} {\bibfnamefont {N.~A.}\ \bibnamefont
  {Sveshnikov}}\ and\ \bibinfo {author} {\bibfnamefont {F.~V.}\ \bibnamefont
  {Tkachov}},\ }\href {\doibase 10.1016/0370-2693(96)00558-8} {\bibfield
  {journal} {\bibinfo  {journal} {Phys. Lett.}\ }\textbf {\bibinfo {volume}
  {B382}},\ \bibinfo {pages} {403} (\bibinfo {year} {1996})},\ \Eprint
  {http://arxiv.org/abs/hep-ph/9512370} {arXiv:hep-ph/9512370 [hep-ph]}
  \BibitemShut {NoStop}%
\bibitem [{\citenamefont {Tkachov}(1997)}]{Tkachov:1995kk}%
  \BibitemOpen
  \bibfield  {author} {\bibinfo {author} {\bibfnamefont {F.~V.}\ \bibnamefont
  {Tkachov}},\ }\href {\doibase 10.1142/S0217751X97002899} {\bibfield
  {journal} {\bibinfo  {journal} {Int. J. Mod. Phys.}\ }\textbf {\bibinfo
  {volume} {A12}},\ \bibinfo {pages} {5411} (\bibinfo {year} {1997})},\ \Eprint
  {http://arxiv.org/abs/hep-ph/9601308} {arXiv:hep-ph/9601308} \BibitemShut
  {NoStop}%
\bibitem [{\citenamefont {Tkachov}(2002)}]{Tkachov:1999py}%
  \BibitemOpen
  \bibfield  {author} {\bibinfo {author} {\bibfnamefont {F.~V.}\ \bibnamefont
  {Tkachov}},\ }\href {\doibase 10.1142/S0217751X02009977} {\bibfield
  {journal} {\bibinfo  {journal} {Int. J. Mod. Phys.}\ }\textbf {\bibinfo
  {volume} {A17}},\ \bibinfo {pages} {2783} (\bibinfo {year} {2002})},\ \Eprint
  {http://arxiv.org/abs/hep-ph/9901444} {arXiv:hep-ph/9901444} \BibitemShut
  {NoStop}%
\bibitem [{\citenamefont {Feige}\ and\ \citenamefont
  {Schwartz}(2014)}]{Feige:2014wja}%
  \BibitemOpen
  \bibfield  {author} {\bibinfo {author} {\bibfnamefont {I.}~\bibnamefont
  {Feige}}\ and\ \bibinfo {author} {\bibfnamefont {M.~D.}\ \bibnamefont
  {Schwartz}},\ }\href {\doibase 10.1103/PhysRevD.90.105020} {\bibfield
  {journal} {\bibinfo  {journal} {Phys. Rev.}\ }\textbf {\bibinfo {volume}
  {D90}},\ \bibinfo {pages} {105020} (\bibinfo {year} {2014})},\ \Eprint
  {http://arxiv.org/abs/1403.6472} {arXiv:1403.6472 [hep-ph]} \BibitemShut
  {NoStop}%
\bibitem [{\citenamefont {Larkoski}\ \emph {et~al.}(2015)\citenamefont
  {Larkoski}, \citenamefont {Neill},\ and\ \citenamefont
  {Stewart}}]{Larkoski:2014bxa}%
  \BibitemOpen
  \bibfield  {author} {\bibinfo {author} {\bibfnamefont {A.~J.}\ \bibnamefont
  {Larkoski}}, \bibinfo {author} {\bibfnamefont {D.}~\bibnamefont {Neill}}, \
  and\ \bibinfo {author} {\bibfnamefont {I.~W.}\ \bibnamefont {Stewart}},\
  }\href {\doibase 10.1007/JHEP06(2015)077} {\bibfield  {journal} {\bibinfo
  {journal} {JHEP}\ }\textbf {\bibinfo {volume} {06}},\ \bibinfo {pages} {077}
  (\bibinfo {year} {2015})},\ \Eprint {http://arxiv.org/abs/1412.3108}
  {arXiv:1412.3108 [hep-th]} \BibitemShut {NoStop}%
\bibitem [{\citenamefont {Breuer}\ and\ \citenamefont
  {Petruccione}(2001)}]{PhysRevA.63.032102}%
  \BibitemOpen
  \bibfield  {author} {\bibinfo {author} {\bibfnamefont {H.-P.}\ \bibnamefont
  {Breuer}}\ and\ \bibinfo {author} {\bibfnamefont {F.}~\bibnamefont
  {Petruccione}},\ }\href {\doibase 10.1103/PhysRevA.63.032102} {\bibfield
  {journal} {\bibinfo  {journal} {Phys. Rev. A}\ }\textbf {\bibinfo {volume}
  {63}},\ \bibinfo {pages} {032102} (\bibinfo {year} {2001})}\BibitemShut
  {NoStop}%
\bibitem [{\citenamefont {Calucci}(2003)}]{PhysRevA.67.042702}%
  \BibitemOpen
  \bibfield  {author} {\bibinfo {author} {\bibfnamefont {G.}~\bibnamefont
  {Calucci}},\ }\href {\doibase 10.1103/PhysRevA.67.042702} {\bibfield
  {journal} {\bibinfo  {journal} {Phys. Rev. A}\ }\textbf {\bibinfo {volume}
  {67}},\ \bibinfo {pages} {042702} (\bibinfo {year} {2003})}\BibitemShut
  {NoStop}%
\bibitem [{\citenamefont {Carney}\ \emph {et~al.}(2017)\citenamefont {Carney},
  \citenamefont {Chaurette}, \citenamefont {Neuenfeld},\ and\ \citenamefont
  {Semenoff}}]{Carney:2017jut}%
  \BibitemOpen
  \bibfield  {author} {\bibinfo {author} {\bibfnamefont {D.}~\bibnamefont
  {Carney}}, \bibinfo {author} {\bibfnamefont {L.}~\bibnamefont {Chaurette}},
  \bibinfo {author} {\bibfnamefont {D.}~\bibnamefont {Neuenfeld}}, \ and\
  \bibinfo {author} {\bibfnamefont {G.~W.}\ \bibnamefont {Semenoff}},\ }\href
  {\doibase 10.1103/PhysRevLett.119.180502} {\bibfield  {journal} {\bibinfo
  {journal} {Phys. Rev. Lett.}\ }\textbf {\bibinfo {volume} {119}},\ \bibinfo
  {pages} {180502} (\bibinfo {year} {2017})},\ \Eprint
  {http://arxiv.org/abs/1706.03782} {arXiv:1706.03782 [hep-th]} \BibitemShut
  {NoStop}%
\bibitem [{\citenamefont {Mehtar-Tani}\ \emph {et~al.}(2013)\citenamefont
  {Mehtar-Tani}, \citenamefont {Milhano},\ and\ \citenamefont
  {Tywoniuk}}]{Mehtar-Tani:2013pia}%
  \BibitemOpen
  \bibfield  {author} {\bibinfo {author} {\bibfnamefont {Y.}~\bibnamefont
  {Mehtar-Tani}}, \bibinfo {author} {\bibfnamefont {J.~G.}\ \bibnamefont
  {Milhano}}, \ and\ \bibinfo {author} {\bibfnamefont {K.}~\bibnamefont
  {Tywoniuk}},\ }\href {\doibase 10.1142/S0217751X13400137} {\bibfield
  {journal} {\bibinfo  {journal} {Int. J. Mod. Phys.}\ }\textbf {\bibinfo
  {volume} {A28}},\ \bibinfo {pages} {1340013} (\bibinfo {year} {2013})},\
  \Eprint {http://arxiv.org/abs/1302.2579} {arXiv:1302.2579 [hep-ph]}
  \BibitemShut {NoStop}%
\bibitem [{\citenamefont {Bauer}\ \emph {et~al.}(2001)\citenamefont {Bauer},
  \citenamefont {Fleming}, \citenamefont {Pirjol},\ and\ \citenamefont
  {Stewart}}]{Bauer:2000yr}%
  \BibitemOpen
  \bibfield  {author} {\bibinfo {author} {\bibfnamefont {C.~W.}\ \bibnamefont
  {Bauer}}, \bibinfo {author} {\bibfnamefont {S.}~\bibnamefont {Fleming}},
  \bibinfo {author} {\bibfnamefont {D.}~\bibnamefont {Pirjol}}, \ and\ \bibinfo
  {author} {\bibfnamefont {I.~W.}\ \bibnamefont {Stewart}},\ }\href {\doibase
  10.1103/PhysRevD.63.114020} {\bibfield  {journal} {\bibinfo  {journal} {Phys.
  Rev.}\ }\textbf {\bibinfo {volume} {D63}},\ \bibinfo {pages} {114020}
  (\bibinfo {year} {2001})},\ \Eprint {http://arxiv.org/abs/hep-ph/0011336}
  {arXiv:hep-ph/0011336} \BibitemShut {NoStop}%
\bibitem [{\citenamefont {Bauer}\ \emph {et~al.}(2002)\citenamefont {Bauer},
  \citenamefont {Pirjol},\ and\ \citenamefont {Stewart}}]{Bauer:2001yt}%
  \BibitemOpen
  \bibfield  {author} {\bibinfo {author} {\bibfnamefont {C.~W.}\ \bibnamefont
  {Bauer}}, \bibinfo {author} {\bibfnamefont {D.}~\bibnamefont {Pirjol}}, \
  and\ \bibinfo {author} {\bibfnamefont {I.~W.}\ \bibnamefont {Stewart}},\
  }\href {\doibase 10.1103/PhysRevD.65.054022} {\bibfield  {journal} {\bibinfo
  {journal} {Phys. Rev.}\ }\textbf {\bibinfo {volume} {D65}},\ \bibinfo {pages}
  {054022} (\bibinfo {year} {2002})},\ \Eprint
  {http://arxiv.org/abs/hep-ph/0109045} {arXiv:hep-ph/0109045} \BibitemShut
  {NoStop}%
\bibitem [{\citenamefont {Bauer}\ and\ \citenamefont
  {Stewart}(2001)}]{Bauer:2001ct}%
  \BibitemOpen
  \bibfield  {author} {\bibinfo {author} {\bibfnamefont {C.~W.}\ \bibnamefont
  {Bauer}}\ and\ \bibinfo {author} {\bibfnamefont {I.~W.}\ \bibnamefont
  {Stewart}},\ }\href {\doibase 10.1016/S0370-2693(01)00902-9} {\bibfield
  {journal} {\bibinfo  {journal} {Phys. Lett.}\ }\textbf {\bibinfo {volume}
  {B516}},\ \bibinfo {pages} {134} (\bibinfo {year} {2001})},\ \Eprint
  {http://arxiv.org/abs/hep-ph/0107001} {arXiv:hep-ph/0107001} \BibitemShut
  {NoStop}%
\bibitem [{\citenamefont {Ellis}\ \emph {et~al.}(2010)\citenamefont {Ellis},
  \citenamefont {Vermilion}, \citenamefont {Walsh}, \citenamefont {Hornig},\
  and\ \citenamefont {Lee}}]{Ellis:2010rwa}%
  \BibitemOpen
  \bibfield  {author} {\bibinfo {author} {\bibfnamefont {S.~D.}\ \bibnamefont
  {Ellis}}, \bibinfo {author} {\bibfnamefont {C.~K.}\ \bibnamefont
  {Vermilion}}, \bibinfo {author} {\bibfnamefont {J.~R.}\ \bibnamefont
  {Walsh}}, \bibinfo {author} {\bibfnamefont {A.}~\bibnamefont {Hornig}}, \
  and\ \bibinfo {author} {\bibfnamefont {C.}~\bibnamefont {Lee}},\ }\href
  {\doibase 10.1007/JHEP11(2010)101} {\bibfield  {journal} {\bibinfo  {journal}
  {JHEP}\ }\textbf {\bibinfo {volume} {11}},\ \bibinfo {pages} {101} (\bibinfo
  {year} {2010})},\ \Eprint {http://arxiv.org/abs/1001.0014} {arXiv:1001.0014
  [hep-ph]} \BibitemShut {NoStop}%
\bibitem [{\citenamefont {Stewart}\ \emph {et~al.}(2010)\citenamefont
  {Stewart}, \citenamefont {Tackmann},\ and\ \citenamefont
  {Waalewijn}}]{Stewart:2010tn}%
  \BibitemOpen
  \bibfield  {author} {\bibinfo {author} {\bibfnamefont {I.~W.}\ \bibnamefont
  {Stewart}}, \bibinfo {author} {\bibfnamefont {F.~J.}\ \bibnamefont
  {Tackmann}}, \ and\ \bibinfo {author} {\bibfnamefont {W.~J.}\ \bibnamefont
  {Waalewijn}},\ }\href {\doibase 10.1103/PhysRevLett.105.092002} {\bibfield
  {journal} {\bibinfo  {journal} {Phys. Rev. Lett.}\ }\textbf {\bibinfo
  {volume} {105}},\ \bibinfo {pages} {092002} (\bibinfo {year} {2010})},\
  \Eprint {http://arxiv.org/abs/1004.2489} {arXiv:1004.2489 [hep-ph]}
  \BibitemShut {NoStop}%
\bibitem [{\citenamefont {Larkoski}\ \emph {et~al.}(2016)\citenamefont
  {Larkoski}, \citenamefont {Moult},\ and\ \citenamefont
  {Neill}}]{Larkoski:2015kga}%
  \BibitemOpen
  \bibfield  {author} {\bibinfo {author} {\bibfnamefont {A.~J.}\ \bibnamefont
  {Larkoski}}, \bibinfo {author} {\bibfnamefont {I.}~\bibnamefont {Moult}}, \
  and\ \bibinfo {author} {\bibfnamefont {D.}~\bibnamefont {Neill}},\ }\href
  {\doibase 10.1007/JHEP05(2016)117} {\bibfield  {journal} {\bibinfo  {journal}
  {JHEP}\ }\textbf {\bibinfo {volume} {05}},\ \bibinfo {pages} {117} (\bibinfo
  {year} {2016})},\ \Eprint {http://arxiv.org/abs/1507.03018} {arXiv:1507.03018
  [hep-ph]} \BibitemShut {NoStop}%
\bibitem [{\citenamefont {Collins}\ and\ \citenamefont
  {Soper}(1981)}]{Collins:1981uk}%
  \BibitemOpen
  \bibfield  {author} {\bibinfo {author} {\bibfnamefont {J.~C.}\ \bibnamefont
  {Collins}}\ and\ \bibinfo {author} {\bibfnamefont {D.~E.}\ \bibnamefont
  {Soper}},\ }\href {\doibase 10.1016/0550-3213(81)90339-4} {\bibfield
  {journal} {\bibinfo  {journal} {Nucl. Phys.}\ }\textbf {\bibinfo {volume}
  {B193}},\ \bibinfo {pages} {381} (\bibinfo {year} {1981})},\ \bibinfo {note}
  {[Erratum: Nucl. Phys.B213,545(1983)]}\BibitemShut {NoStop}%
\bibitem [{\citenamefont {Kidonakis}\ \emph {et~al.}(1998)\citenamefont
  {Kidonakis}, \citenamefont {Oderda},\ and\ \citenamefont
  {Sterman}}]{Kidonakis:1998nf}%
  \BibitemOpen
  \bibfield  {author} {\bibinfo {author} {\bibfnamefont {N.}~\bibnamefont
  {Kidonakis}}, \bibinfo {author} {\bibfnamefont {G.}~\bibnamefont {Oderda}}, \
  and\ \bibinfo {author} {\bibfnamefont {G.~F.}\ \bibnamefont {Sterman}},\
  }\href {\doibase 10.1016/S0550-3213(98)00441-6} {\bibfield  {journal}
  {\bibinfo  {journal} {Nucl. Phys.}\ }\textbf {\bibinfo {volume} {B531}},\
  \bibinfo {pages} {365} (\bibinfo {year} {1998})},\ \Eprint
  {http://arxiv.org/abs/hep-ph/9803241} {arXiv:hep-ph/9803241} \BibitemShut
  {NoStop}%
\bibitem [{\citenamefont {Breuer}\ and\ \citenamefont
  {Petruccione}(2002)}]{OpenSysBook}%
  \BibitemOpen
  \bibfield  {author} {\bibinfo {author} {\bibfnamefont {H.-P.}\ \bibnamefont
  {Breuer}}\ and\ \bibinfo {author} {\bibfnamefont {F.}~\bibnamefont
  {Petruccione}},\ }\href@noop {} {\emph {\bibinfo {title} {The Theory of Open
  Quantum Systems}}}\ (\bibinfo  {publisher} {Oxford University Press},\
  \bibinfo {year} {2002})\BibitemShut {NoStop}%
\bibitem [{\citenamefont {Bassetto}\ \emph {et~al.}(1982)\citenamefont
  {Bassetto}, \citenamefont {Ciafaloni}, \citenamefont {Marchesini},\ and\
  \citenamefont {Mueller}}]{Bassetto:1982ma}%
  \BibitemOpen
  \bibfield  {author} {\bibinfo {author} {\bibfnamefont {A.}~\bibnamefont
  {Bassetto}}, \bibinfo {author} {\bibfnamefont {M.}~\bibnamefont {Ciafaloni}},
  \bibinfo {author} {\bibfnamefont {G.}~\bibnamefont {Marchesini}}, \ and\
  \bibinfo {author} {\bibfnamefont {A.~H.}\ \bibnamefont {Mueller}},\ }\href
  {\doibase 10.1016/0550-3213(82)90161-4} {\bibfield  {journal} {\bibinfo
  {journal} {Nucl. Phys.}\ }\textbf {\bibinfo {volume} {B207}},\ \bibinfo
  {pages} {189} (\bibinfo {year} {1982})}\BibitemShut {NoStop}%
\bibitem [{\citenamefont {Bassetto}\ \emph {et~al.}(1983)\citenamefont
  {Bassetto}, \citenamefont {Ciafaloni},\ and\ \citenamefont
  {Marchesini}}]{Bassetto:1984ik}%
  \BibitemOpen
  \bibfield  {author} {\bibinfo {author} {\bibfnamefont {A.}~\bibnamefont
  {Bassetto}}, \bibinfo {author} {\bibfnamefont {M.}~\bibnamefont {Ciafaloni}},
  \ and\ \bibinfo {author} {\bibfnamefont {G.}~\bibnamefont {Marchesini}},\
  }\href {\doibase 10.1016/0370-1573(83)90083-2} {\bibfield  {journal}
  {\bibinfo  {journal} {Phys. Rept.}\ }\textbf {\bibinfo {volume} {100}},\
  \bibinfo {pages} {201} (\bibinfo {year} {1983})}\BibitemShut {NoStop}%
\bibitem [{\citenamefont {Mueller}(1983)}]{Mueller:1982cq}%
  \BibitemOpen
  \bibfield  {author} {\bibinfo {author} {\bibfnamefont {A.~H.}\ \bibnamefont
  {Mueller}},\ }\href {\doibase 10.1016/0550-3213(83)90176-1} {\bibfield
  {journal} {\bibinfo  {journal} {Nucl. Phys.}\ }\textbf {\bibinfo {volume}
  {B213}},\ \bibinfo {pages} {85} (\bibinfo {year} {1983})}\BibitemShut
  {NoStop}%
\bibitem [{\citenamefont {Catani}\ \emph {et~al.}(1992)\citenamefont {Catani},
  \citenamefont {Dokshitzer}, \citenamefont {Fiorani},\ and\ \citenamefont
  {Webber}}]{Catani:1991pm}%
  \BibitemOpen
  \bibfield  {author} {\bibinfo {author} {\bibfnamefont {S.}~\bibnamefont
  {Catani}}, \bibinfo {author} {\bibfnamefont {Y.~L.}\ \bibnamefont
  {Dokshitzer}}, \bibinfo {author} {\bibfnamefont {F.}~\bibnamefont {Fiorani}},
  \ and\ \bibinfo {author} {\bibfnamefont {B.~R.}\ \bibnamefont {Webber}},\
  }\href {\doibase 10.1016/0550-3213(92)90296-N} {\bibfield  {journal}
  {\bibinfo  {journal} {Nucl. Phys.}\ }\textbf {\bibinfo {volume} {B377}},\
  \bibinfo {pages} {445} (\bibinfo {year} {1992})}\BibitemShut {NoStop}%
\bibitem [{\citenamefont {Khoze}\ \emph {et~al.}(2000)\citenamefont {Khoze},
  \citenamefont {Ochs},\ and\ \citenamefont {Wosiek}}]{Khoze:2000iq}%
  \BibitemOpen
  \bibfield  {author} {\bibinfo {author} {\bibfnamefont {V.~A.}\ \bibnamefont
  {Khoze}}, \bibinfo {author} {\bibfnamefont {W.}~\bibnamefont {Ochs}}, \ and\
  \bibinfo {author} {\bibfnamefont {J.}~\bibnamefont {Wosiek}},\ }\href@noop {}
  {\  (\bibinfo {year} {2000})},\ \Eprint {http://arxiv.org/abs/hep-ph/0009298}
  {arXiv:hep-ph/0009298} \BibitemShut {NoStop}%
\bibitem [{\citenamefont {Gerwick}\ \emph {et~al.}(2013)\citenamefont
  {Gerwick}, \citenamefont {Schumann}, \citenamefont {Gripaios},\ and\
  \citenamefont {Webber}}]{Gerwick:2012fw}%
  \BibitemOpen
  \bibfield  {author} {\bibinfo {author} {\bibfnamefont {E.}~\bibnamefont
  {Gerwick}}, \bibinfo {author} {\bibfnamefont {S.}~\bibnamefont {Schumann}},
  \bibinfo {author} {\bibfnamefont {B.}~\bibnamefont {Gripaios}}, \ and\
  \bibinfo {author} {\bibfnamefont {B.}~\bibnamefont {Webber}},\ }\href
  {\doibase 10.1007/JHEP04(2013)089} {\bibfield  {journal} {\bibinfo  {journal}
  {JHEP}\ }\textbf {\bibinfo {volume} {04}},\ \bibinfo {pages} {089} (\bibinfo
  {year} {2013})},\ \Eprint {http://arxiv.org/abs/1212.5235} {arXiv:1212.5235
  [hep-ph]} \BibitemShut {NoStop}%
\bibitem [{\citenamefont {Elder}\ \emph {et~al.}(2017)\citenamefont {Elder},
  \citenamefont {Procura}, \citenamefont {Thaler}, \citenamefont {Waalewijn},\
  and\ \citenamefont {Zhou}}]{Elder:2017bkd}%
  \BibitemOpen
  \bibfield  {author} {\bibinfo {author} {\bibfnamefont {B.~T.}\ \bibnamefont
  {Elder}}, \bibinfo {author} {\bibfnamefont {M.}~\bibnamefont {Procura}},
  \bibinfo {author} {\bibfnamefont {J.}~\bibnamefont {Thaler}}, \bibinfo
  {author} {\bibfnamefont {W.~J.}\ \bibnamefont {Waalewijn}}, \ and\ \bibinfo
  {author} {\bibfnamefont {K.}~\bibnamefont {Zhou}},\ }\href {\doibase
  10.1007/JHEP06(2017)085} {\bibfield  {journal} {\bibinfo  {journal} {JHEP}\
  }\textbf {\bibinfo {volume} {06}},\ \bibinfo {pages} {085} (\bibinfo {year}
  {2017})},\ \Eprint {http://arxiv.org/abs/1704.05456} {arXiv:1704.05456
  [hep-ph]} \BibitemShut {NoStop}%
\bibitem [{\citenamefont {Andersson}\ \emph {et~al.}(1988)\citenamefont
  {Andersson}, \citenamefont {Dahlkvist},\ and\ \citenamefont
  {Gustafson}}]{Andersson:1988ee}%
  \BibitemOpen
  \bibfield  {author} {\bibinfo {author} {\bibfnamefont {B.}~\bibnamefont
  {Andersson}}, \bibinfo {author} {\bibfnamefont {P.}~\bibnamefont
  {Dahlkvist}}, \ and\ \bibinfo {author} {\bibfnamefont {G.}~\bibnamefont
  {Gustafson}},\ }\href {\doibase 10.1016/0370-2693(88)90128-1} {\bibfield
  {journal} {\bibinfo  {journal} {Phys. Lett.}\ }\textbf {\bibinfo {volume}
  {B214}},\ \bibinfo {pages} {604} (\bibinfo {year} {1988})}\BibitemShut
  {NoStop}%
\bibitem [{\citenamefont {Andersson}\ \emph {et~al.}(1989)\citenamefont
  {Andersson}, \citenamefont {Dahlqvist},\ and\ \citenamefont
  {Gustafson}}]{Andersson:1989ww}%
  \BibitemOpen
  \bibfield  {author} {\bibinfo {author} {\bibfnamefont {B.}~\bibnamefont
  {Andersson}}, \bibinfo {author} {\bibfnamefont {P.}~\bibnamefont
  {Dahlqvist}}, \ and\ \bibinfo {author} {\bibfnamefont {G.}~\bibnamefont
  {Gustafson}},\ }\href {\doibase 10.1007/BF01415560} {\bibfield  {journal}
  {\bibinfo  {journal} {Z. Phys.}\ }\textbf {\bibinfo {volume} {C44}},\
  \bibinfo {pages} {455} (\bibinfo {year} {1989})}\BibitemShut {NoStop}%
\bibitem [{\citenamefont {Dahlqvist}\ \emph {et~al.}(1989)\citenamefont
  {Dahlqvist}, \citenamefont {Andersson},\ and\ \citenamefont
  {Gustafson}}]{Dahlqvist:1989yc}%
  \BibitemOpen
  \bibfield  {author} {\bibinfo {author} {\bibfnamefont {P.}~\bibnamefont
  {Dahlqvist}}, \bibinfo {author} {\bibfnamefont {B.}~\bibnamefont
  {Andersson}}, \ and\ \bibinfo {author} {\bibfnamefont {G.}~\bibnamefont
  {Gustafson}},\ }\href {\doibase 10.1016/0550-3213(89)90092-8} {\bibfield
  {journal} {\bibinfo  {journal} {Nucl. Phys.}\ }\textbf {\bibinfo {volume}
  {B328}},\ \bibinfo {pages} {76} (\bibinfo {year} {1989})}\BibitemShut
  {NoStop}%
\bibitem [{\citenamefont {Gustafson}\ and\ \citenamefont
  {Nilsson}(1991)}]{Gustafson:1991ru}%
  \BibitemOpen
  \bibfield  {author} {\bibinfo {author} {\bibfnamefont {G.}~\bibnamefont
  {Gustafson}}\ and\ \bibinfo {author} {\bibfnamefont {A.}~\bibnamefont
  {Nilsson}},\ }\href {\doibase 10.1007/BF01559451} {\bibfield  {journal}
  {\bibinfo  {journal} {Z. Phys.}\ }\textbf {\bibinfo {volume} {C52}},\
  \bibinfo {pages} {533} (\bibinfo {year} {1991})}\BibitemShut {NoStop}%
\bibitem [{\citenamefont {Davighi}\ and\ \citenamefont
  {Harris}(2018)}]{Davighi:2017hok}%
  \BibitemOpen
  \bibfield  {author} {\bibinfo {author} {\bibfnamefont {J.}~\bibnamefont
  {Davighi}}\ and\ \bibinfo {author} {\bibfnamefont {P.}~\bibnamefont
  {Harris}},\ }\href {\doibase 10.1140/epjc/s10052-018-5819-8} {\bibfield
  {journal} {\bibinfo  {journal} {Eur. Phys. J.}\ }\textbf {\bibinfo {volume}
  {C78}},\ \bibinfo {pages} {334} (\bibinfo {year} {2018})},\ \Eprint
  {http://arxiv.org/abs/1703.00914} {arXiv:1703.00914 [hep-ph]} \BibitemShut
  {NoStop}%
\bibitem [{\citenamefont {Dreyer}\ \emph {et~al.}(2018)\citenamefont {Dreyer},
  \citenamefont {Salam},\ and\ \citenamefont {Soyez}}]{Dreyer:2018nbf}%
  \BibitemOpen
  \bibfield  {author} {\bibinfo {author} {\bibfnamefont {F.~A.}\ \bibnamefont
  {Dreyer}}, \bibinfo {author} {\bibfnamefont {G.~P.}\ \bibnamefont {Salam}}, \
  and\ \bibinfo {author} {\bibfnamefont {G.}~\bibnamefont {Soyez}},\
  }\href@noop {} {\  (\bibinfo {year} {2018})},\ \Eprint
  {http://arxiv.org/abs/1807.04758} {arXiv:1807.04758 [hep-ph]} \BibitemShut
  {NoStop}%
\bibitem [{\citenamefont {Chien}\ and\ \citenamefont
  {Kunnawalkam~Elayavalli}(2018)}]{Chien:2018dfn}%
  \BibitemOpen
  \bibfield  {author} {\bibinfo {author} {\bibfnamefont {Y.-T.}\ \bibnamefont
  {Chien}}\ and\ \bibinfo {author} {\bibfnamefont {R.}~\bibnamefont
  {Kunnawalkam~Elayavalli}},\ }\href@noop {} {\  (\bibinfo {year} {2018})},\
  \Eprint {http://arxiv.org/abs/1803.03589} {arXiv:1803.03589 [hep-ph]}
  \BibitemShut {NoStop}%
\bibitem [{\citenamefont {Komiske}\ \emph {et~al.}(2018)\citenamefont
  {Komiske}, \citenamefont {Metodiev},\ and\ \citenamefont
  {Thaler}}]{Komiske:2017aww}%
  \BibitemOpen
  \bibfield  {author} {\bibinfo {author} {\bibfnamefont {P.~T.}\ \bibnamefont
  {Komiske}}, \bibinfo {author} {\bibfnamefont {E.~M.}\ \bibnamefont
  {Metodiev}}, \ and\ \bibinfo {author} {\bibfnamefont {J.}~\bibnamefont
  {Thaler}},\ }\href {\doibase 10.1007/JHEP04(2018)013} {\bibfield  {journal}
  {\bibinfo  {journal} {JHEP}\ }\textbf {\bibinfo {volume} {04}},\ \bibinfo
  {pages} {013} (\bibinfo {year} {2018})},\ \Eprint
  {http://arxiv.org/abs/1712.07124} {arXiv:1712.07124 [hep-ph]} \BibitemShut
  {NoStop}%
\bibitem [{\citenamefont {Cogan}\ \emph {et~al.}(2015)\citenamefont {Cogan},
  \citenamefont {Kagan}, \citenamefont {Strauss},\ and\ \citenamefont
  {Schwarztman}}]{Cogan:2014oua}%
  \BibitemOpen
  \bibfield  {author} {\bibinfo {author} {\bibfnamefont {J.}~\bibnamefont
  {Cogan}}, \bibinfo {author} {\bibfnamefont {M.}~\bibnamefont {Kagan}},
  \bibinfo {author} {\bibfnamefont {E.}~\bibnamefont {Strauss}}, \ and\
  \bibinfo {author} {\bibfnamefont {A.}~\bibnamefont {Schwarztman}},\ }\href
  {\doibase 10.1007/JHEP02(2015)118} {\bibfield  {journal} {\bibinfo  {journal}
  {JHEP}\ }\textbf {\bibinfo {volume} {02}},\ \bibinfo {pages} {118} (\bibinfo
  {year} {2015})},\ \Eprint {http://arxiv.org/abs/1407.5675} {arXiv:1407.5675
  [hep-ph]} \BibitemShut {NoStop}%
\bibitem [{\citenamefont {de~Oliveira}\ \emph {et~al.}(2016)\citenamefont
  {de~Oliveira}, \citenamefont {Kagan}, \citenamefont {Mackey}, \citenamefont
  {Nachman},\ and\ \citenamefont {Schwartzman}}]{deOliveira:2015xxd}%
  \BibitemOpen
  \bibfield  {author} {\bibinfo {author} {\bibfnamefont {L.}~\bibnamefont
  {de~Oliveira}}, \bibinfo {author} {\bibfnamefont {M.}~\bibnamefont {Kagan}},
  \bibinfo {author} {\bibfnamefont {L.}~\bibnamefont {Mackey}}, \bibinfo
  {author} {\bibfnamefont {B.}~\bibnamefont {Nachman}}, \ and\ \bibinfo
  {author} {\bibfnamefont {A.}~\bibnamefont {Schwartzman}},\ }\href {\doibase
  10.1007/JHEP07(2016)069} {\bibfield  {journal} {\bibinfo  {journal} {JHEP}\
  }\textbf {\bibinfo {volume} {07}},\ \bibinfo {pages} {069} (\bibinfo {year}
  {2016})},\ \Eprint {http://arxiv.org/abs/1511.05190} {arXiv:1511.05190
  [hep-ph]} \BibitemShut {NoStop}%
\bibitem [{\citenamefont {Baldi}\ \emph {et~al.}(2016)\citenamefont {Baldi},
  \citenamefont {Bauer}, \citenamefont {Eng}, \citenamefont {Sadowski},\ and\
  \citenamefont {Whiteson}}]{Baldi:2016fql}%
  \BibitemOpen
  \bibfield  {author} {\bibinfo {author} {\bibfnamefont {P.}~\bibnamefont
  {Baldi}}, \bibinfo {author} {\bibfnamefont {K.}~\bibnamefont {Bauer}},
  \bibinfo {author} {\bibfnamefont {C.}~\bibnamefont {Eng}}, \bibinfo {author}
  {\bibfnamefont {P.}~\bibnamefont {Sadowski}}, \ and\ \bibinfo {author}
  {\bibfnamefont {D.}~\bibnamefont {Whiteson}},\ }\href {\doibase
  10.1103/PhysRevD.93.094034} {\bibfield  {journal} {\bibinfo  {journal} {Phys.
  Rev.}\ }\textbf {\bibinfo {volume} {D93}},\ \bibinfo {pages} {094034}
  (\bibinfo {year} {2016})},\ \Eprint {http://arxiv.org/abs/1603.09349}
  {arXiv:1603.09349 [hep-ex]} \BibitemShut {NoStop}%
\bibitem [{\citenamefont {Guest}\ \emph {et~al.}(2016)\citenamefont {Guest},
  \citenamefont {Collado}, \citenamefont {Baldi}, \citenamefont {Hsu},
  \citenamefont {Urban},\ and\ \citenamefont {Whiteson}}]{Guest:2016iqz}%
  \BibitemOpen
  \bibfield  {author} {\bibinfo {author} {\bibfnamefont {D.}~\bibnamefont
  {Guest}}, \bibinfo {author} {\bibfnamefont {J.}~\bibnamefont {Collado}},
  \bibinfo {author} {\bibfnamefont {P.}~\bibnamefont {Baldi}}, \bibinfo
  {author} {\bibfnamefont {S.-C.}\ \bibnamefont {Hsu}}, \bibinfo {author}
  {\bibfnamefont {G.}~\bibnamefont {Urban}}, \ and\ \bibinfo {author}
  {\bibfnamefont {D.}~\bibnamefont {Whiteson}},\ }\href {\doibase
  10.1103/PhysRevD.94.112002} {\bibfield  {journal} {\bibinfo  {journal} {Phys.
  Rev.}\ }\textbf {\bibinfo {volume} {D94}},\ \bibinfo {pages} {112002}
  (\bibinfo {year} {2016})},\ \Eprint {http://arxiv.org/abs/1607.08633}
  {arXiv:1607.08633 [hep-ex]} \BibitemShut {NoStop}%
\end{thebibliography}%

\end{document}